# Synthesis of Quantum Vector Databases Based on Grover's Algorithm

**Cesar Borisovich Pronin**

**Andrey Vladimirovich Ostroukh**

MOSCOW AUTOMOBILE AND ROAD CONSTRUCTION STATE TECHNICAL UNIVERSITY (MADI)., 64, Leningradsky prospect, Moscow, Russia

**Abstract:** This paper describes a method for using Grover's algorithm to create a quantum vector database, the database stores embeddings based on Controlled-S gates, which represent a binary numerical value. This value represents the embedding's value. The process of creating meaningful embeddings is handled by a classical computer and the search process is handled by the quantum computer. This search approach might be beneficial for a large enough database, or it could be seen as a very qubit-efficient (super dense) way for storing data on a quantum computer, since the proposed circuit stores many embeddings inside one quantum register simultaneously.

## Introduction

Vector databases offer efficient data processing, improved data organization and visualization, better scalability, and in some cases their implementation and use isn't much more complex compared to regular databases. They are particularly well-suited for Artificial Intelligence (AI) related applications involving similarity search, machine learning, and natural language processing. By leveraging vector similarity search algorithms, vector databases enable faster and more accurate querying, leading to improved performance and accuracy in AI systems [1-4].

Right now, vector databases are often used for information retrieval tasks on their own, or assisting Large Language Models. Vector databases provide valuable context while preserving context length by filtering information irrelevant to the user's query. This is important, since context length is rather limited in even the most advanced and popular modern AI models, like GPT-4 and LLaMA.

## Models and methods

Grover's quantum search algorithm could be used in various ways to speed up search problems by querying all possible states of a quantum register in parallel, something that could be impractical or impossible using classical computational methods [5-7]. Its computational complexity of Grover's algorithm is $O\left(\sqrt{\frac{N}{M}}\right)$, where $N$ is the number of states of a quantum register and $M$ is the number of awaited solutions [8]. The proposed circuit is based on Grover's algorithm and, since during search in our oracle two similar Controlled-S gates together will form a Controlled-Z gate – the complexity of the proposed method is the same as the complexity of Grover's algorithm. Grover's algorithm has an amplitude amplification step (also known as Grover diffusion operator), that amplifies the amplitudes of states selected by the oracle (fig. 1).

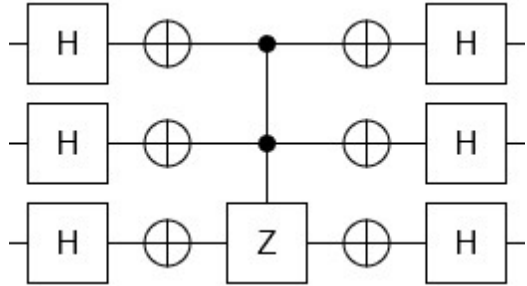

*Fig. 1. The amplitude amplification step of Grover's algorithm that is composed into the "Amp" gate*

The common implementation of Grover's algorithm, described by IBM, uses one (or a set of) Controlled Pauli-Z gate inside the oracle function, their goal is to invert the amplitude sign of the target states (fig. 2). This implementation of Grover's algorithm has an interesting feature that in most cases could be seen as a limiting factor – when an oracle returns a certain number of solutions, their amplitude can no longer be amplified, thus these solutions will have the same probability of occurring after measurement as the other states (fig. 3). This makes the search result non-definitive. Based on our observations, the maximum number of unique solutions that could be coded by CZ gates in the oracle of Grover's algorithm of "n" qubits is $N_{sol} = 2^{n-1} - 1$.

The coding of a state by a gate in these quantum circuits done like this: the "1" in the binary number is a control, the "0" is the anti-control (control surrounded by NOT gates), the controlled gate itself is treated as a "1", unless it is inverted by NOT gates surrounding it.

From the above description, one approach is to create a database composed of CZ gates using this principle, with $N_{sol} + 1$ as the number of states coded inside the database (if the number of input states is less, the database could be padded to $N_{sol} +$ 1 length). Then the query would be coded as a similar CZ gate and if it exists in the database it would just disrupt the system (by utilizing the reversibility principle of QC) making the probability of other states coded in such database more prominent after measurement. This approach has some apparent drawbacks: it could be queried by only one state at a time to be definitive, could require a high number of Grover iterations to be definitive, needs to be padded or limited by $N_{sol}$ states e.t.c.

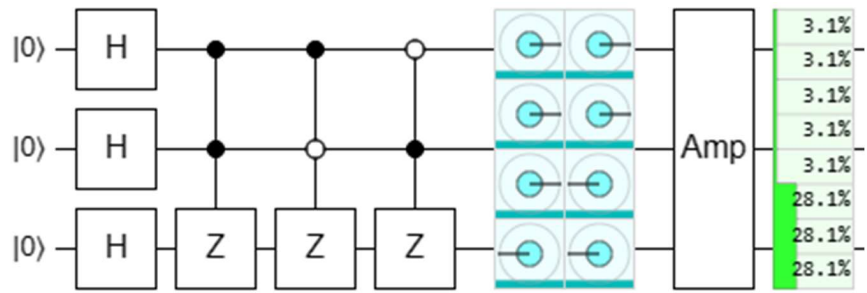

*Fig. 2. Using Controlled Pauli-Z gates to code $N_{sol}$ number of states, coded states are 111, 101, 110 (gates ordered from left to right)*

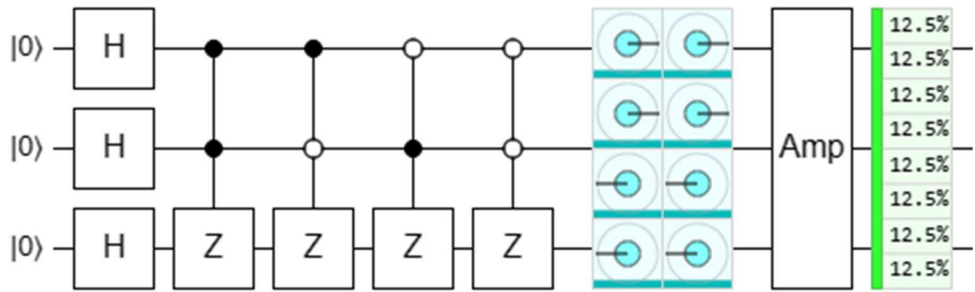

*Fig. 3. Using Controlled Pauli-Z gates to code $N_{sol} + 1$ number of states, coded states are 111, 101, 110, 100 (gates ordered from left to right)*

We propose a vector database that is formed by using a series of controlled S gates. The query – an input vector or embedding, is also coded as a corresponding controlled S gate.

1) If the input vector value exists in the database – this value will have an increased probability at the output of the algorithm (fig. 4).
2) If the value doesn't exist in the database – this value's probability won't be amplified, compared to the probability of other embeddings contained in the database (fig. 5).

It is possible to search for a number values at the same time (fig. 6)

This approach, although different, was inspired by the image coding method in [9], the possibility of creating quantum vector databases was mentioned here [10]. A number of researchers have proposed other fascinating methods of performing search, using quantum computing [11-14]. Quantum simulator "Quirk" was used to visualize quantum circuits [15] and IBM Quantum [16] and the Qiskit framework [17] were used for testing the circuits on real quantum computers.

The ideal situation for this type of circuit is achieved when the total number of vectors in a query plus the database is equal to $N_{sol} + 2$ (fig. 7). But this isn't a limiting factor, because in other cases:

1) If the input embedding exists in the database – the measurement probability of the input embedding's state will be significantly increased compared to other states (fig. 8).
2) If the input embedding doesn't exist in the database - the amplitude of the input embedding's probability will be similar to other states in the quantum register (fig. 9).

Of course, this search approach isn't flawless, the increased amplitude values of target states could "blend in" with amplitude values of other states coded in the database. This behavior could be rectified depending on the situation - in some cases by using more qubits, or, in other cases, by adding "padding" embeddings to the database to reach an optimal state for comparing measurement results before and after the query. These false embeddings could be marked as such on a classical computer during oracle\embedding preparation to remove them from the search results.

Summarized – an input embedding's amplitude increases significantly compared to all other states if it exists in the database. This corresponds in it being returned after measurement significantly more often. On fig 4-9 gates "INPUT" and "DB" are labels added for clarity, they don't affect the circuit.

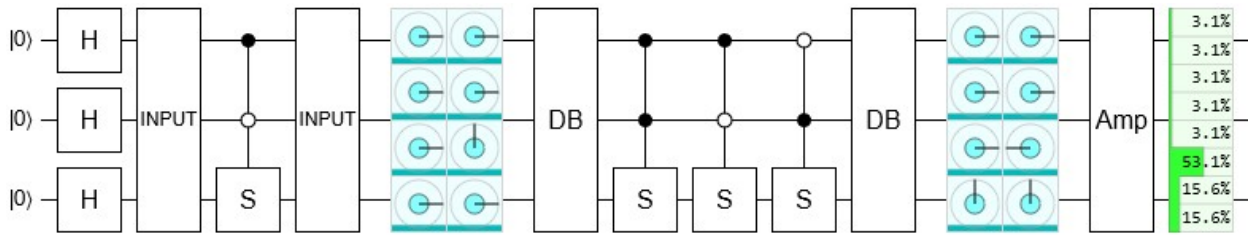

*Fig. 4. Searching for an existing item (binary value 101) in the proposed quantum vector database, state 101 has a significantly higher probability of being returned after measurement*

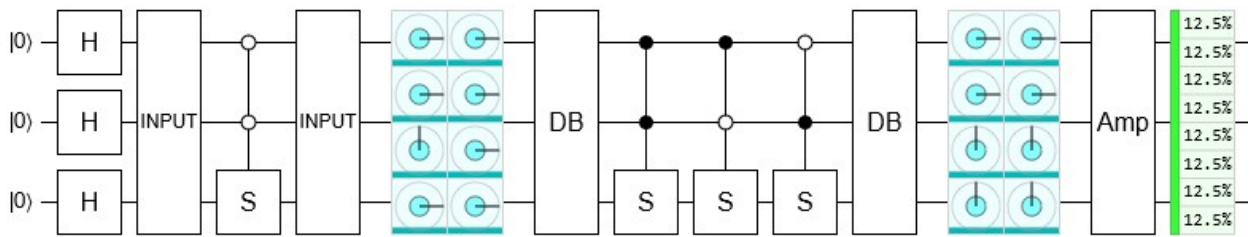

*Fig. 5. Searching for a non-existent item (binary value 100) in the proposed quantum vector database, all states have equal probability of being returned after measurement*

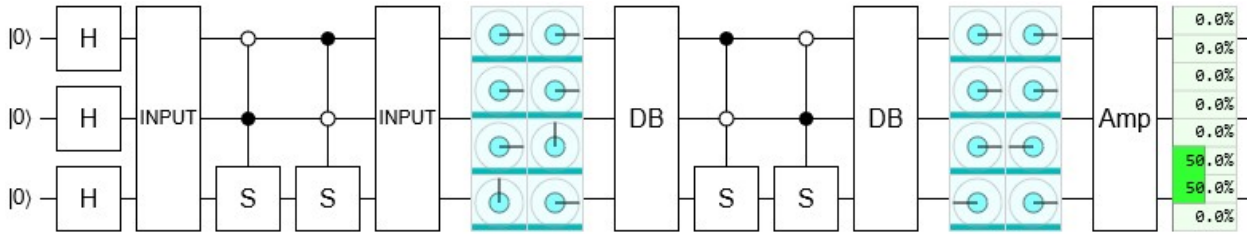

*Fig. 6. Attempt to search for two values at a time*

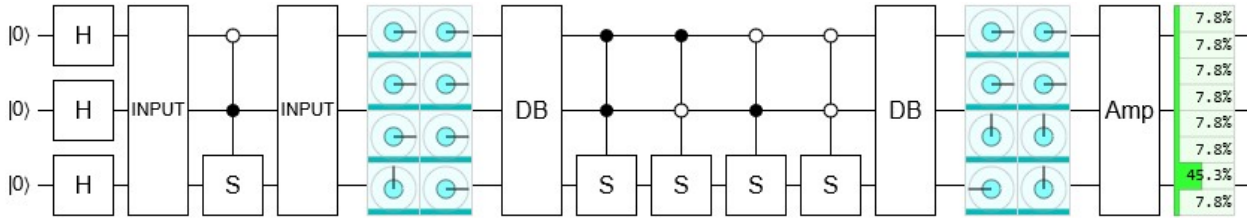

*Fig. 7. Ideal situation for CS-gate based search with Grover's algorithm*

Another benefit of using Controlled-S gates in the oracle makes it possible to store more values in the database than $N_{sol} + 1$. Let's call this approach "overprovisioning" (fig. 8 - 9).

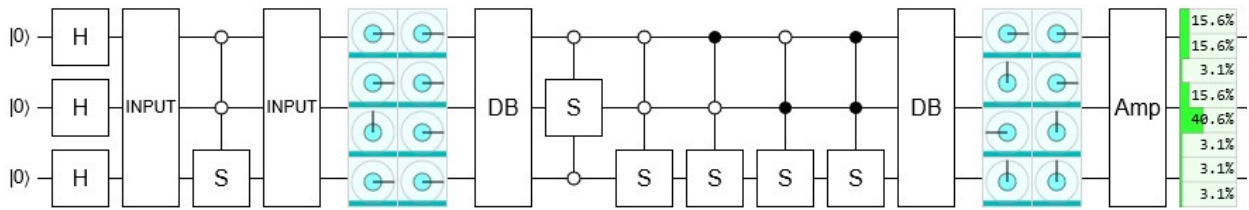

*Fig. 8. Attempt to search an "overprovisioned" quantum database for an existing value*

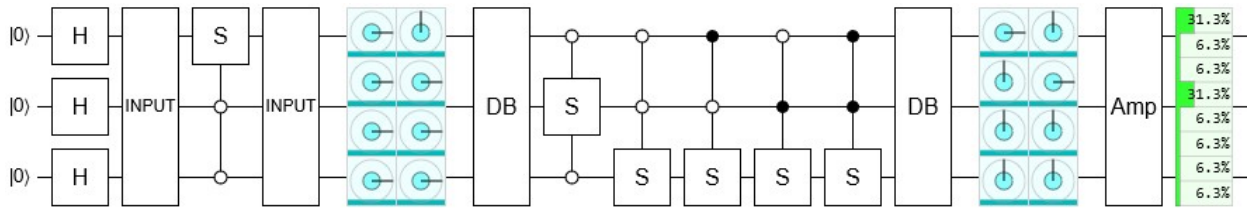

*Fig. 9. Attempt to search an "overprovisioned" quantum database for a non-existent value*

## Results

One of the key benefits of our proposed approach for creating quantum vector databases is that demonstration circuits are quite simple and require a low amount of qubits. This makes it feasible to launch these circuits on real universal quantum computers. The oracle part that contains the embedding database could be also composed in a form of a single matrix that is the multiplication product of the matrices for Controlled-S gates that make up the database.

The proof-of-concept algorithm on fig. 4-5 is small enough to be tested on a real quantum computer. We used Qiskit to synthesize QASM code for the circuit [18] in this example, mostly because S-gates with multiple controls aren't readily available in IBM Quantum Composer and need to be synthesized. Another benefit of synthesis is the ability to easily resize the circuit to a required number of qubits. In the given example we used IBM-Lagos with 7 qubits and quantum "volume" of 32, both test circuits ran with 4096 shots (fig. 10-11).

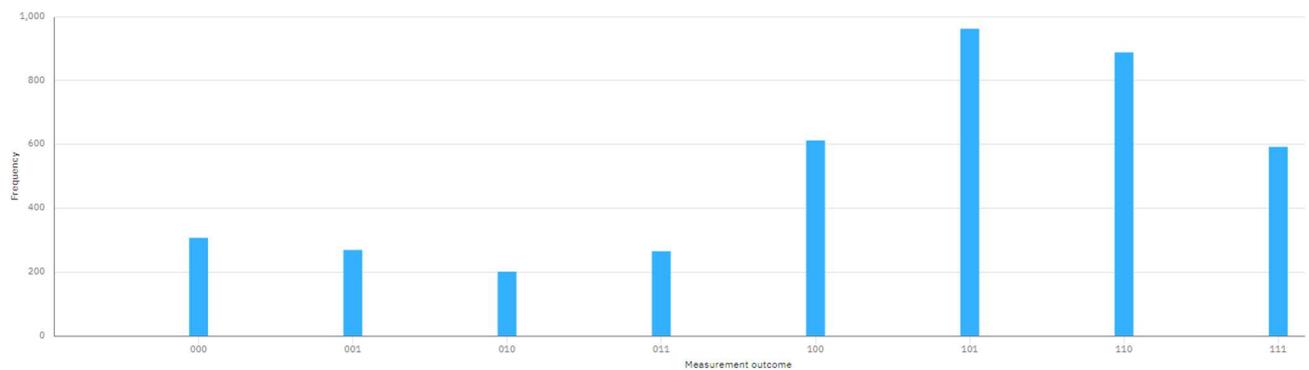

*Fig. 10. The result of testing the circuit from fig. 4 on IBM's quantum computer (search for existing value)*

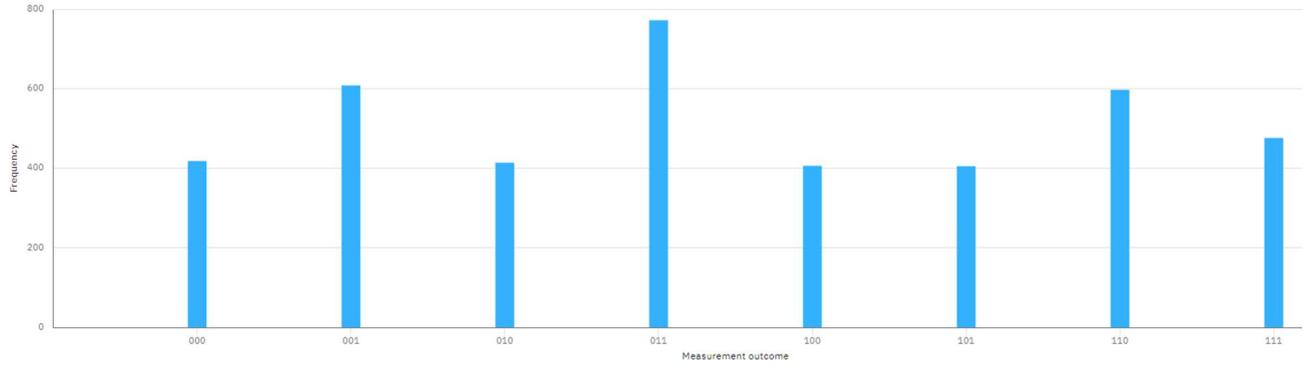

*Fig. 11. The result of testing the circuit from fig. 5 on IBM's quantum computer (search for non-existant value)*

## Conclusion

In this paper an approach of creating quantum vector databases based on Controlled-S gates and Grover's algorithm was proposed, the results were tested both on a quantum simulator and a universal quantum computer. The proposed circuits are available on GitHub [18].

We consider that this approach has the potential to speed up the process of querying large databases, even with varied data. This could be useful in many fields that require analyzing a large array of statistical data. In the realm of AI, it is beneficial to have a database of specific information assisting a large language model in information retrieval and analysis tasks, since this eliminates the need for the language model to perceive a huge context of unrelated data and significantly decreases the chance of generating incorrect or misleading outputs. Having a huge database could decrease performance on a classical system, requiring faster storage solutions and significant processing power for indexing, this is where a quantum vector database could provide speedup and potential decrease of energy consumption.

# References

1) Reimers N, Gurevych I. Sentence-bert: Sentence embeddings using siamese bert-networks. arXiv preprint arXiv:1908.10084. 2019 Aug 27.

2) Thakur N, Reimers N, Daxenberger J, Gurevych I. Augmented sbert: Data augmentation method for improving bi-encoders for pairwise sentence scoring tasks. arXiv preprint arXiv:2010.08240. 2020 Oct 16.

3) SBERT. Retrieve & Re-Rank. Retrieved from https://sbert.net/examples/applications/retrieve_rerank/README.html

4) Malkov YA, Yashunin DA. Efficient and robust approximate nearest neighbor search using hierarchical navigable small world graphs. IEEE transactions on pattern analysis and machine intelligence. 2018 Dec 28;42(4):824-36.

5) Ostroukh A.V., Pronin C.B.: Researching the Possibilities of Creating Mathematical Oracle Functions for Grover's Quantum Search Algorithm (in Russian) // Industrial Automatic Control Systems and Controllers. - 2018. - No. 9. - P. 3-10. URL: http://asu.tgizd.ru/

6) Pronin CB, Ostroukh AV. Developing Mathematical Oracle Functions for Grover Quantum Search Algorithm. arXiv preprint arXiv:2109.05921. 2021 Sep 3.

7) Grover LK. A fast quantum mechanical algorithm for database search. InProceedings of the twenty-eighth annual ACM symposium on Theory of computing 1996 Jul 1 (pp. 212-219).

8) Ulyanov SV, Ulyanov VS. Quantum Algorithmic Gate-Based Computing: Grover Quantum Search Algorithm Design in Quantum Software Engineering. arXiv preprint arXiv:2304.13703. 2023 Apr 20.

9) Tezuka H, Nakaji K, Satoh T, Yamamoto N. Grover search revisited: Application to image pattern matching. Physical Review A. 2022 Mar 24;105(3):032440.

10) Chakraborty, Sayantan (2022). A Prototype For Quantum Database in Hybrid Quantum. TechRxiv. Preprint. https://doi.org/10.36227/techrxiv.20237202.v6

11) Schuld M, Killoran N. Quantum machine learning in feature hilbert spaces. Physical review letters. 2019 Feb 1;122(4):040504.

12) Mani A, Patvardhan C. A fast measurement based fixed-point quantum search algorithm. InInternational Conference on Quantum Information 2011 Jun 6 (p. QMI28). Optica Publishing Group.

13) Mani A, Patvardhan C. A Fast fixed-point Quantum Search Algorithm by using Disentanglement and Measurement. arXiv preprint arXiv:1203.3178. 2012 Mar 14.

14) Yuan G, Lu J, Su P. Quantum-Inspired Keyword Search on Multi-model Databases. InDatabase Systems for Advanced Applications: 26th International Conference, DASFAA 2021, Taipei, Taiwan, April 11–14, 2021, Proceedings, Part II 26 2021 (pp. 585-602). Springer International Publishing.

15) Quirk // Quirk – online quantum computer simulator. URL: http://algassert.com/quirk

16) IBM Quantum Computing – 2023, URL: https://www.ibm.com/quantum .

17) Qiskit: An Open-source Framework for Quantum Computing - 2023, doi: 10.5281/zenodo.2573505

18) Pronin C.B. Quantum vector database based on Grover's algorithm. GitHub: https://github.com/MexIvanov/QuantumVectorDB

**Author details**

**Andrey Vladimirovich Ostroukh**, Russian Federation, full member RAE, Doctor of Technical Sciences, Professor, Department «Automated Control Systems». State Technical University – MADI, 125319, Russian Federation, Moscow, Leningradsky prospekt, 64. Tel.: +7 (499) 151-64-12. http://www.madi.ru , email: ostroukh@mail.ru , ORCID: https://orcid.org/0000-0002-8887-6132

**Cesar Borisovich Pronin**, Russian Federation, Postgraduate Student, Department «Automated Control Systems». State Technical University – MADI, 125319, Russian Federation, Moscow, Leningradsky prospekt, 64. Tel.: +7 (499) 151-64-12. http://www.madi.ru , email: caesarpr12@gmail.com , ORCID: https://orcid.org/0000-0002-9994-1032